\documentclass[epjST]{svjour}
\usepackage{graphics}
\usepackage{amsmath}
\usepackage{amsfonts}

\newcommand{\ket}[1]{\vert #1 \rangle}
\newcommand{\bra}[1]{\langle #1 \vert}

\newcommand{\braket}[2]{\langle #1 \vert #2 \rangle}
\newcommand{\meanvalue}[3]{\langle #1 \vert #2 \vert #3 \rangle}
\newcommand{\ketbra}[2]{\vert #1 \rangle \langle #2 \vert}

\newcommand{\imm}{{\rm i }}

\begin{document}
\title{The coherent interaction between matter and radiation}
\subtitle{A tutorial on the Jaynes-Cummings model}
\author{Matteo Bina\thanks{\email{matteo.bina@gmail.com}}}
\institute{Dipartimento di Scienza e Alta Tecnologia, Universit\`{a} degli Studi dell'Insubria, Via Valleggio 11, I-22100 Como, Italy.}
\abstract{
The Jaynes-Cummings (JC) model is a milestone in the theory of coherent interaction between a two-level system and a single bosonic field mode. This tutorial aims to give a complete description of the model, analyzing the Hamiltonian of the system, its eigenvalues and eigestates, in order to characterize the dynamics of system and subsystems. The Rabi oscillations, together with the collapse and revival effects, are distinguishing features of the JC model and are important for applications in Quantum Information theory. The framework of cavity quantum electrodynamics (cQED) is chosen and two fundamental experiments on the coherent interaction between Rydberg atoms and a single cavity field mode are described.
} 
\maketitle

\section{Introduction}
The JC model was originally proposed in 1963 by Edwin Jaynes and Fred Cummings \cite{JC} in order to study the relationship between the quantum theory of radiation and the semi-classical theory in describing the phenomenon of spontaneous emission. In the semi-classical theory of atom-field interaction, the field is treated as a definite function of time rather than as an operator, whereas the atom is quantized. The semi-classical theory can explain many phenomena that are observed in modern optics, for example the existence of Rabi cycles in atomic excitation probabilities for radiation fields with sharply defined energy (narrow bandwidth). The JC model aims to find how quantization of the radiation field affects the predictions for the evolution of the state of a two-level system, in comparison with semi-classical theory of matter-radiation interaction. What is a purely quantum effect, described only by the JC model and not by the semi-classical theory, is the revival of the atomic population inversion after its collapse, providing a direct evidence of the discreteness of field states (photons).

The JC model is still an important and actual topic in quantum physics, since it touches different branches of the most recent and outstanding research activities, and big efforts have been carried out to extend this basic theoretical topic \cite{Shor_Knight}. The natural framework for the implementation and experimental testing of the JC model is cQED, where fundamental experiments has been performed using microwave and optical cavities \cite{Rempe1987,Brune_Haroche,Hagley_Haroche,Mabuchi_optical}. The JC model has been also used in the framework of trapped ions, where the stable elctronic states of the ions interact with their vibrational modes (phonons) \cite{Trapped_Ions}. In the context of solid state systems, semiconductor quantum dots are placed inside photonic crystal, micropillar or microdisk resonators \cite{Solid_State}. Nowadays the advanced technology in circuit-QED allows to build a two-level system from quantum circuits interacting with transmission line microwave resonators \cite{Circuit_QED}.

In order to more precisely describe the interaction between an atom and a laser field, the model is generalized in different ways. Some of the generalizations comprehend different initial conditions \cite{JCExt_InitConditions}, dissipation and damping in the model \cite{JCExt_Dissipation}, multilevel atoms and multiple atoms \cite{JCExt_multilevel}, and multi-mode description of the field \cite{JCExt_multimode}. Other extensions of the JC model include a driving pump laser acting on one or more two-level atoms, which allows the simultaneous action of rotating and counter-rotating coupling terms in the Hamiltonian, including dissipative effects due to Markovian environments \cite{Solano_Walther,Lougovski_Casagrande,Bina_Casagrande,Bina_Casagrande_EPJD}.

The JC model serves as a tight connection between Quantum Optics and Quantum Information Theory, since it naturally provides the dynamics of interacting qubits, the basic amount of quantum information encoded in every two-level system. Several JC-based systems have been employed in order to perform quantum computation algorithms, quantum teleportation and quantum cryptography protocols \cite{Nielsen_Chuang}.

The present tutorial is structured in a pedagogic form, in order to provide all the mathematical passages necessary to the comprehension of the basic JC model and its figures of merit. The physical model is introduced in Sec. 1, then the uncoupled and coupled systems are analyzed respectively in Sec. 2 and Sec. 3. The diagonalization is performed in Sec. 4, bringing to the complete diagram of the energy levels of the system. Sec. 5 is devoted to the description of the dyanamics, with a particular stress on the resonant case, where the transition frequency of the two-level atom is equal to that of the cavity field mode. A formalization of the JC model is performed in Sec. 6, by introducing the density operator of the system and of the subsystems. Vacuum Rabi oscillations and phenomena of collapses and revivals are the objects of, respectively, Sec. 7 and Sec. 8. The final part of the tutorial is devoted to illustrate two model experiments in cQED, where the quantum nature of matter-radiation interaction shows itself in a clear and powerful way.

\section{The model}
The fully quantized JC model concerns a two-level system interacting with one radiation mode. Under the dipole approximation and the rotating wave approximation (RWA), this kind of system is described by the following Hamiltonian:
\begin{equation}\label{HJC}
\mathcal{H}=\mathcal{H}_0+\mathcal{H}_{AF}=\frac{\hbar\omega_A}{2}r_3+\hbar\omega(a^\dag a +\frac{1}{2})+\hbar g(ar^++a^\dag r^-)
\end{equation}
where $\mathcal{H}_{AF}=\hbar g(ar^++a^\dag r^-)$ is the interaction term and $\mathcal{H}_0$ is the free Hamiltonian.
The atomic transition frequency is $\omega_A$, the frequency associated to the field mode is $\omega$, while the coupling constant $g$ quantifies the strength of the interaction between the atom and the electromagnetic field. The atomic \emph{excited state} is denoted as $\ket{e}$, with energy $E_e=\hbar\omega_A/2$ and the \emph{ground state} as $\ket{g}$, with energy $E_g=-\hbar\omega_A/2$. The atomic \emph{raising operator} is $r^+=\ketbra{e}{g}$, the atomic \emph{lowering operator} is $r^-=\ketbra{g}{e}$ and $r_3=\ketbra{e}{e}-\ketbra{g}{g}$ is called \emph{inversion operator}. The atomic transition energy associated the inversion operator $r_3$ is $\hbar\omega_A=E_e-E_g$. The field mode \emph{creation operator} is $a^\dag$, whereas the field mode \emph{annihilation operator} is $a$. \\
The operators introduced above satisfy the commutation relations 
\begin{equation}\begin{split}
[a,a^\dag]&=\mathbb{I}\\
[r^+,r^-]&=r_3\\
[r_3,r^\pm]&=\pm 2 r^\pm
\end{split}\end{equation}
providing the following conditions for the Hamiltonian operators:\\

\begin{itemize}
\item \emph{at resonance} $\omega=\omega_A$ $$[\mathcal{H}_0,\mathcal{H}_{AF}]=0\text{ and also }[\mathcal{H}_0,\mathcal{H}]=0$$
meaning that $\mathcal{H}_0$ and $\mathcal{H}$ have common eigenvalues;\\

\item the total number of excitation quanta is conserved: $[N,\mathcal{H}]=0$, where $N\equiv a^\dag a+ r^+r^-$ is a constant of motion.
\end{itemize}

\subsection{Rotating wave approximation (RWA)}
It must be noted that the original and general Hamiltonian describing the interaction of a two-level system with a quantized field mode, is different from Eq. (\ref{HJC}) and includes the so called anti-rotating terms, $r^+a^\dag$ and $r^-a$. In fact, under the dipole approximation, the two-level atom can be described by a dipole moment $\hat{d}$ coupled to an electromagnetic field $\hat{E}\propto (a+a^\dag)$ and the interaction reads 
\begin{equation}
\mathcal{H}_{1}=-\hat{d}\cdot\hat{E}.
\end{equation}
The atomic dipole moment can be expressed in terms of the atomic basis states and only the off-diagonal terms are non-zero since, by parity considerations, $\meanvalue{e}{\hat{d}}{e}=\meanvalue{g}{\hat{d}}{g}=0$ and one can write
\begin{equation}
\hat{d}=d\ketbra{g}{e}+d^*\ketbra{e}{g}=d(r^-+r^+)
\end{equation}
where $\meanvalue{e}{d}{g}=d$ and it has been assumed, without loss of generality, that $d$ is real. Hence, the interaction between the two-level atom and the quantized field mode is
\begin{equation}
\mathcal{H}_{1}=\hbar g (r^++r^-)(a^\dag+a)
\end{equation}
where the coupling constant $g$ depends on the strength of the interaction between the atomic dipole and the field mode. The Hamiltonian describing the whole system is then $\mathcal{H}=\mathcal{H}_{0}+\mathcal{H}_{1}$. This is the general way to describe the quantized matter-radiation interaction, which goes under the name of Rabi model.\\
In order to recover the Hamiltonian (\ref{HJC}) describing the fully quantized JC model, the RWA is employed to discard the anti-rotating terms, $r^+a^\dag$ and $r^-a$. This fact is appreciated if the system dynamics is described in the interaction picture, considering the transformation $\ket{\Psi(t)}_I=U_t\ket{\Psi(t)}$ with $U_t={\rm exp}(\frac{\imm}{\hbar}\mathcal{H}_0t)$ and where the state vector $\ket{\Psi(t)}$ is a solution of the Schr\"odinger equation
\begin{equation}
\imm\hbar\frac{\partial}{\partial t}\ket{\Psi(t)}=\mathcal{H}\ket{\Psi(t)}.
\end{equation}
A similar expression can be derived for the state vector $\ket{\Psi(t)}_I$ in the interaction picture
\begin{equation}
\imm\hbar\frac{\partial}{\partial t}\ket{\Psi(t)}_I=\mathcal{H}_I(t)\ket{\Psi(t)}_I
\end{equation}
with 
\begin{equation}
\mathcal{H}_I(t)=U_t^\dag\mathcal{H}_1U_t=\hbar g \left ( r^+a^\dag{\rm e}^{\imm(\omega+\omega_A)t}+ra{\rm e}^{-\imm(\omega+\omega_A)t}+r^+a{\rm e}^{-\imm(\omega-\omega_A)t}+ra^\dag{\rm e}^{\imm(\omega-\omega_A)t} \right)
\end{equation}
Close to the resonance condition, $\omega\simeq\omega_A$, the terms $r^+a^\dag$ and $ra$ are multiplied by the factors ${\rm e}^{\pm(\omega+\omega_A)t}$ which are fast rotating in time. The RWA allows to neglect these terms in describing the dynamics of the system and to recover the JC Hamiltonian (\ref{HJC}), since 
\begin{equation}
\mathcal{H}_I(t)\simeq\hbar g \left (r^+a{\rm e}^{-\imm(\omega-\omega_A)t}+ra^\dag{\rm e}^{\imm(\omega-\omega_A)t} \right)
\end{equation}
which gives in the Schr\"odinger picture the typical JC interacting term $\mathcal{H}_{AF}$.\\
A more mathematical approach for the RWA is to consider the Dyson series expansion of the time evolution operator in the interaction picture
\begin{equation}
U_I(t)=\mathbb{I}-\frac{\imm}{\hbar}\int_0^t\mathcal{H}_I(t_1)dt_1+\left (- \frac{\imm}{\hbar} \right)^2 \int_0^tdt_1\int_0^{t_1}\mathcal{H}_I(t_1)\mathcal{H}_I(t_2)dt_2+\ldots
\end{equation}
which rules the time evolution of the state vector in the interaction picture
\begin{equation}
\ket{\Psi(t)}_I=U_I(t)\ket{\Psi(0)}_I.
\end{equation}
Considering, for instance, the first order in the series expansion, the integration provides terms as $\pm(\omega+\omega_A)^{-1}$ connected to the operators $r^+a^\dag$ and $ra$, whereas terms like $\pm(\omega-\omega_A)^{-1}$ are linked to $r^+a$ and $ra^\dag$. The last terms are dominant with respect to the first ones under the resonance condition $\omega\simeq\omega_A$, together with the assumption that $\left \{ \frac{g}{\omega},\frac{g}{\omega_A} \right \}\ll 1$. With these conditions, the major contributions to the time evolution operator come from the resonant terms $r^+a$ and $ra^\dag$.\\
The JC model is valid only if the RWA can be employed to describe an effective coherent interaction between a two-level system and a field mode. This can be done whenever the coupling constant $g$ is much smaller than the single frequencies $\omega$ and $\omega_A$.
Recently, new regimes has been explored both theoretically \cite{USC_theo} and experimentally \cite{USC_exp} where the coupling constant $g$ is comparable with the atomic and field frequencies, for which the RWA is no longer applicable.

\section{Uncoupled system}
Without any interaction between the atom and the cavity field mode ($g=0$), the Hamiltonian of the uncoupled system is
\begin{equation}\label{H0}
\mathcal{H}=\mathcal{H}_{0}=\hbar\left [ \frac{\omega_A}{2}r_3+\omega \left ( a^\dag a+\frac{1}{2} \right ) \right ].
\end{equation}
One can distinguish two main situations, the \emph{resonant} and the \emph{non resonant} case:\\

\begin{figure}[!h]
\centering
\resizebox{1\columnwidth}{!}{\includegraphics{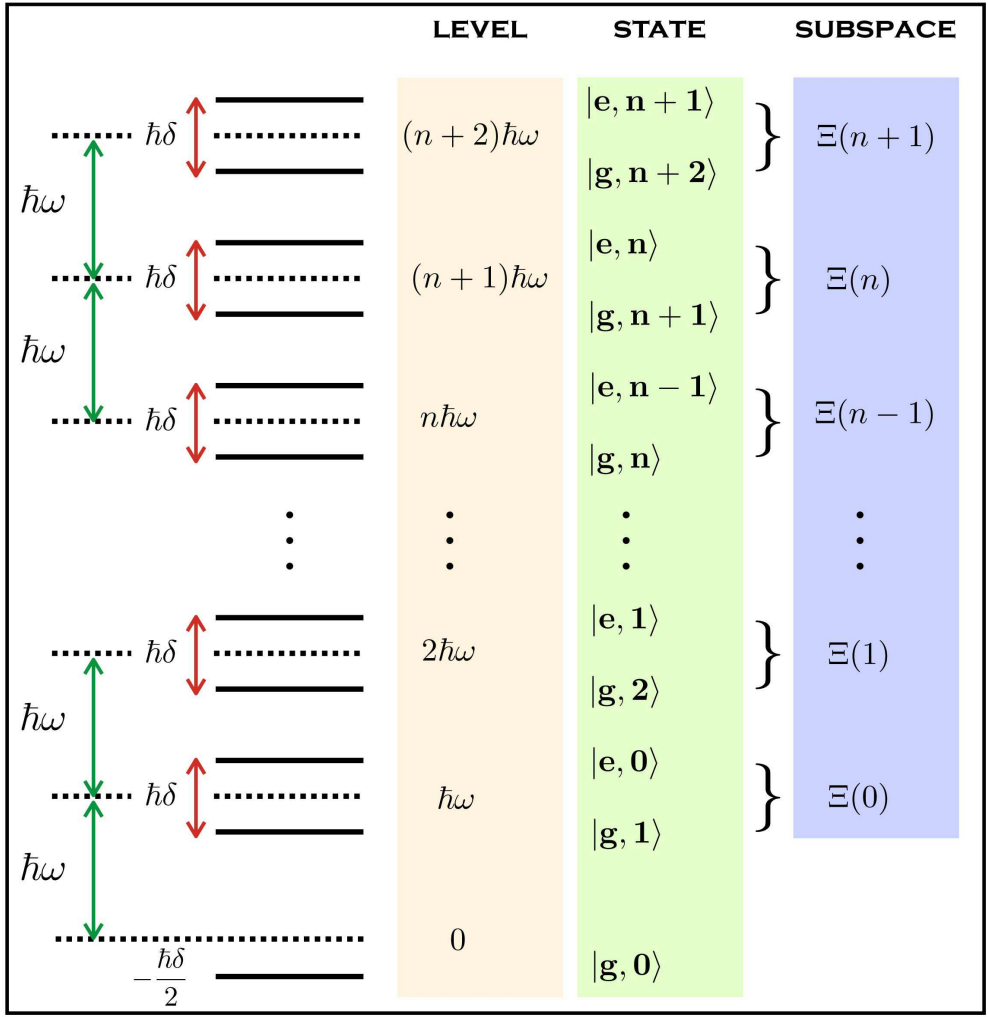}}
\caption{\label{NakedLevels}Scheme of the naked levels $E_{e,n}$ and $E_{g,n+1}$ of the uncoupled system described by $\mathcal{H}_0$. Each couple of energy levels corresponds to a couple of vector eigenstates, which form a bidimensional Hilbert subspace $\Xi(n)$. The ground state $\ket{g,0}$ is independent from the others.}
\end{figure}

\begin{itemize}
\item At resonance ($\omega=\omega_A$) Eq. \ref{H0} becomes
\begin{equation}
\mathcal{H}_0=\hbar\omega \left ( \frac{r_3}{2}+a^\dag a +\frac{1}{2} \right)=\hbar\omega (r^+r^-+a^\dag a)=\hbar\omega N
\end{equation}
Finding eigenstates and eigenvalues of $\mathcal{H}_0$ is a quite easy task. Referring to the previous observation about the conservation of the number of quanta of the system, one can consider the system states $\ket{e,n}\equiv\ket{e}\otimes\ket{n}$ (atom in the excited state and $n$ photons) and $\ket{g,n+1}\equiv\ket{g}\otimes\ket{n+1}$ (atom in the ground state and $n+1$ photons), where $\ket{n}$ (with $n=0,1,\ldots$) denotes the Fock basis states of a bosonic field mode. These states are eigenstates of the number operator $N$ with the same eigenvalue $n+1$ and, hence, of the system Hamiltonian $\mathcal{H}_0$ with the eigenvalue $\hbar\omega(n+1)$. The only exception is the ground state of the whole system $\ket{g,0}$, for which $\mathcal{H}_0\ket{g,0}=0$ holds.\\

\item In general, for a non-resonant condition, the Hamiltonian of the uncoupled system can be written as
\begin{equation}
\mathcal{H}_0=\omega \left ( \frac{r_3}{2}+a^\dag a +\frac{1}{2} \right)+ \frac{\hbar\delta}{2}r_3
\end{equation}
where the \emph{detuning parameter} $\delta\equiv\omega_A-\omega$ has been introduced. Now, as depicted in Fig. \ref{NakedLevels}, the two system eigenstates $\ket{e,n}$ and $\ket{g,n+1}$ have different eigenvalues
\begin{subequations}\begin{align}
\mathcal{H}_0\ket{e,n}&=E_{e,n}\ket{e,n}=\hbar\left [ (n+1)\omega+\frac{\delta}{2} \right ]\ket{e,n}\\
\mathcal{H}_0\ket{g,n+1}&=E_{g,n+1}\ket{g,n+1}=\hbar\left [ (n+1)\omega-\frac{\delta}{2} \right ]\ket{g,n+1}
\end{align}\end{subequations}
giving rise to $n$ orthogonal Hilbert subspaces $\Xi(n)$ of dimension 2x2. 
\end{itemize}

\section{Coupled system}
Introducing the coupling term in the Hamiltonian of the system, $\mathcal{H}_{AF}=\hbar g (a r^++a^\dag r^-)$, the previous separated systems (atom and field mode) become a unique entangled system. One can notice immediately that $\mathcal{H}_{AF}$ couples only those states that belong to a single subspace (or bidimensional manifold) $\Xi(n)$. In fact:
\begin{itemize}
\item the process of \emph{absorption} is mathematically described by the matrix element $$\meanvalue{e,n}{\mathcal{H}_{AF}}{g,n+1}=\hbar g\sqrt{n+1},$$ which corresponds to the transition $\ket{g,n+1}\rightarrow\ket{e,n}$;
\item the \emph{emission} process is given by $$\meanvalue{g,n+1}{\mathcal{H}_{AF}}{e,n}=\hbar g\sqrt{n+1},$$ which corresponds to the transition $\ket{e,n}\rightarrow\ket{g,n+1}$.
\end{itemize}
Hence, it is sufficient to consider only the bidimensional manifold, eigenspace of $\mathcal{H}_0$ at a given number of photons $n$:
\begin{equation}
\Xi(n) : \{ \ket{e,n},\ket{g,n+1} \} \quad (n=0,1,2,\ldots).
\end{equation}
Within each of these subspaces, the Hamiltonian of the system is given by the matrix
\begin{equation}\begin{split}
(\mathcal{H})_n&=
\begin{pmatrix}
\meanvalue{g,n+1}{\mathcal{H}_{0}}{g,n+1} & \meanvalue{g,n+1}{\mathcal{H}_{AF}}{e,n}\\[1.5ex]
\meanvalue{e,n}{\mathcal{H}_{AF}}{g,n+1} & \meanvalue{e,n}{\mathcal{H}_{0}}{e,n}
\end{pmatrix}\\
&=\hbar
\begin{pmatrix}
(n+1)\omega +\frac{\delta}{2} & g\sqrt{n+1}\\[1.5ex]
 g\sqrt{n+1} & (n+1)\omega -\frac{\delta}{2}
\end{pmatrix}
\end{split}\end{equation}
which can be rewritten as 
\begin{equation}\label{HforDiag}
(\mathcal{H})_n=\hbar\omega(n+1)\begin{pmatrix} 1 & 0 \\ 0 & 1
\end{pmatrix}
+\frac{\hbar}{2}\begin{pmatrix} \delta & \Omega_n \\ \Omega_n & -\delta
\end{pmatrix}
\end{equation}
where $\Omega_n\equiv 2g\sqrt{n+1}$ is called \emph{Rabi frequency} for $n$ photons.

\section{Diagonalization}
In this section the eigenvalues and eigenvectors of Eq. (\ref{HforDiag}) are calculated.
\subsection{Eigenvalues}
\begin{figure}[!h]
\centering
\resizebox{1\columnwidth}{!}{\includegraphics{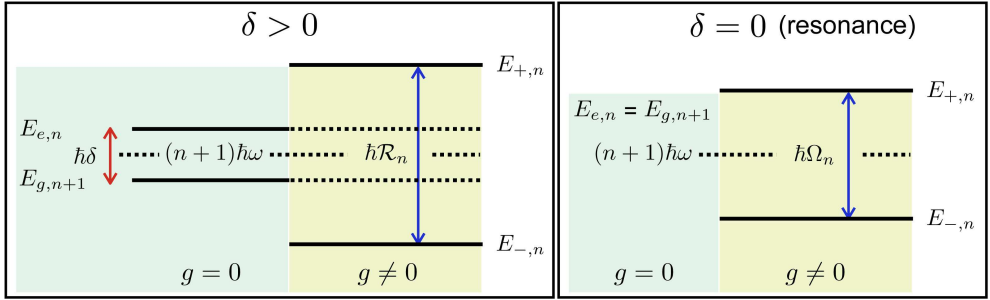}}
\caption{\label{DressedLevels}Scheme of the dressed levels $E_{\pm n}$ of the Hamiltonian $\mathcal{H}$ of the coupled system, related to that of the uncoupled system. At resonance $\delta=0$ the degeneracy is lifted by the coupling and the separation between the two energy levels is $(\Delta E)_n=\hbar\Omega_n$.}
\end{figure}

By imposing the eigenvalue equation
\begin{equation*}
\hbar
\begin{pmatrix}
(n+1)\omega +\frac{\delta}{2} & \frac{\Omega_n}{2}\\[1.5ex]
\frac{\Omega_n}{2} & (n+1)\omega -\frac{\delta}{2}
\end{pmatrix}
\begin{pmatrix} u \\ v \end{pmatrix}
=\hbar\lambda\begin{pmatrix} u \\ v \end{pmatrix}
\end{equation*}
one obtains the two energy eigenvalues:
\begin{equation}\label{eigenvalues}
E_{\pm,n}=\hbar\lambda_{\pm}=\hbar\left [ (n+1)\omega\pm\frac{\mathcal{R}_n}{2}\right ]
\end{equation}
where $\mathcal{R}_n=\sqrt{\Omega_n^2+\delta^2}$ is called \emph{generalized Rabi frequency} from which, at resonance conditions $\delta=0$, one gets the Rabi frequency $\mathcal{R}_n=\Omega_n$.
At resonance the degeneracy of the two energy states of the system is lifted and the energy separation is given by $(\Delta E)_n=\hbar\Omega_n$. In this sense the Rabi frequency plays the role of a Bohr frequency for the whole system and the energy separation of the two levels is higher as the number of photons increases, $(\Delta E)_n\propto\sqrt{n}$ (see Fig. \ref{DressedLevels}).

\subsection{Eigenstates}
\begin{figure}[!h]
\centering
\resizebox{0.65\columnwidth}{!}{\includegraphics{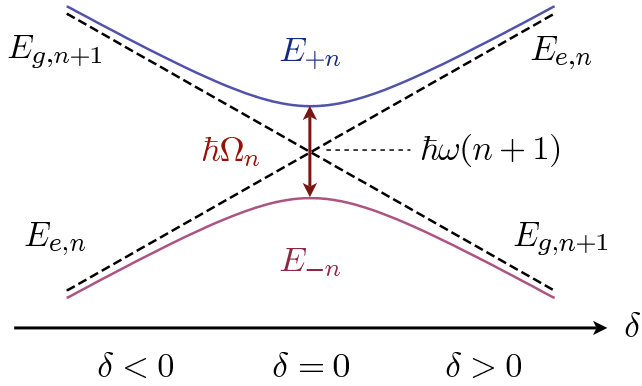}}
\caption{\label{LevelDiagram}Dressed eigenvalues in function of $\delta=\omega_A-\omega$. The coupling removes the degeneracy at $\delta=0$, whereas the naked eigenvalues cross each other. When $\delta\gg 1$ the dressed eigenvalues tend to the naked ones, as the atomic and field subsystems were uncoupled.}
\end{figure}

The eigenstates of the system corresponding to the eigenvalues (\ref{eigenvalues}) are also called \emph{dressed states}, as the ``naked'' atomic states are now ``dressed'' by means of the interaction with photons. By imposing the eigenvalue equation in matrix form (\ref{HforDiag}) one obtains
\begin{equation}
\begin{pmatrix} \delta & \Omega_n\\ 
\Omega_n & -\delta \end{pmatrix} \begin{pmatrix} u \\ v \end{pmatrix}
=\pm\mathcal{R}_n \begin{pmatrix} u \\ v \end{pmatrix}
\end{equation}
and after some passages, using the normalization condition $u^2+v^2=1$, the eigenvectors are found to be:
\begin{subequations}\begin{align}
&\begin{pmatrix} u \\ v \end{pmatrix}_+=\frac{1}{\sqrt{(\mathcal{R}_n-\delta)^2+\Omega_n^2}}\begin{pmatrix} \Omega_n \\ \mathcal{R}_n-\delta \end{pmatrix}\\
&\begin{pmatrix} u \\ v \end{pmatrix}_-=\frac{1}{\sqrt{(\mathcal{R}_n-\delta)^2+\Omega_n^2}}\begin{pmatrix}  \mathcal{R}_n-\delta \\ -\Omega_n \end{pmatrix}.
\end{align}\end{subequations}
The two eigenvectors can be rewritten in a more compact form, in terms of basis eigenvectors, as:
\begin{subequations}\begin{align}
&\begin{pmatrix} u \\ v \end{pmatrix}_+=\sin\theta_n\begin{pmatrix} 1 \\ 0 \end{pmatrix}+\cos\theta_n\begin{pmatrix} 0 \\ 1 \end{pmatrix}\\
&\begin{pmatrix} u \\ v \end{pmatrix}_+=\cos\theta_n\begin{pmatrix} 1 \\ 0 \end{pmatrix}-\sin\theta_n\begin{pmatrix} 0 \\ 1 \end{pmatrix}
\end{align}\end{subequations}
where $\sin\theta_n\equiv\frac{\Omega_n}{\sqrt{(\mathcal{R}_n-\delta)^2+\Omega_n^2}}$, $\cos\theta_n\equiv\frac{\mathcal{R}_n-\delta}{\sqrt{(\mathcal{R}_n-\delta)^2+\Omega_n^2}}$ and $\tan 2\theta_n=-\frac{\Omega_n}{\delta}$.\\
Going back from the eigenvectors (matrix form) to the state vectors (Dirac notation) within the manifold $\Xi(n)$, one has:
\begin{equation}
\mathcal{H}\ket{\pm n}=E_{\pm,n}\ket{\pm n}=\hbar\left[ (n+1)\omega\pm\frac{\mathcal{R}_n}{2} \right ] \ket{\pm n}
\end{equation}
where the dressed states in fucntion of the naked states are
\begin{equation}\begin{split}
\ket{+n}&=\sin\theta_n\ket{e,n}+\cos\theta_n\ket{g,n+1}\\
\ket{-n}&=\cos\theta_n\ket{e,n}-\sin\theta_n\ket{g,n+1}
\end{split}\end{equation}
which are general entangled states of the global system atom-field.\\
The interaction Hamiltonian $\mathcal{H}_{AF}$ produces a unitary transformation on the basis states $\{\ket{e,n},\ket{g,n+1}\}$, which is simply a rotation of an angle $\theta_n$ in $\Xi(n)$. In particular, at resonance condition one obtains:
\begin{equation}\label{max_ent_states}
\ket{\pm n}=\frac{1}{\sqrt{2}} \left ( \ket{e,n}\pm\ket{g,n+1} \right )
\end{equation}
which is a maximally entangled state, with $\theta_n=\frac{\pi}{4}$. From the relation $\tan2\theta_n=-\frac{\Omega_n}{\delta}$ one gets the states (\ref{max_ent_states}) as $\delta\rightarrow 0$, while, if $|\delta|\rightarrow\infty$ the mixing angle $\theta_n\rightarrow 0$ and the uncoupled (or naked) states are recovered.\\
Within each subspace $\Xi(n)$ and varying the detuning parameter $\delta$, the corresponding levels' chart is depicted in Fig. \ref{LevelDiagram}.

\subsubsection{Naked levels} 
 The eigenvalues of $\mathcal{H}_0$ are linear in $\delta$:
\begin{align*}
E_{e,n}&=\hbar\left [ (n+1)\omega+\frac{\delta}{2} \right ]\\
E_{g,n+1}&=\hbar\left [ (n+1)\omega-\frac{\delta}{2} \right ]
\end{align*}
and they cross at $\delta=0$ showing a degeneracy at resonance $E_{e,n}=E_{g,n+1}=(n+1)\hbar\omega$.
\subsubsection{Dressed levels}
As obtained before, the eigenvalues of $\mathcal{H}$ are $E_{\pm n}\hbar\left [ (n+1)\omega\pm\frac{\mathcal{R}_n}{2} \right ]$, with $\mathcal{R}_n\equiv\sqrt{\Omega_n^2+\delta^2}$.\\
At resonance ($\delta=0$) the energy levels do not cross, but they are separated by the effect of the interaction by the quantity $(\Delta E)_n=\hbar\Omega_n=2\hbar g \sqrt{n+1}$.\\
On the other hand, quite off resonance $|\delta|\gg \Omega_n$, the generalized Rabi frequency is $\mathcal{R}_n\simeq|\delta|$ and the dressed states are asymptotically equal to the naked states. This phisically means that the eigenstates of the total Hamiltonian $\mathcal{H}$ coincide with those of the uncoupled Hamiltonian $\mathcal{H}_0$ or, equivalently, it means that $g\rightarrow 0$.

\section{Dynamics}
In this section the dynamics of an initial pure state under the action of the Hamiltonian (\ref{HJC}) is considered. In the Schr\"odinger picture the evolved state is given by
\begin{equation}\label{Psi_t}
\ket{\Psi(t)}={\rm e}^{-\frac{\imm}{\hbar}\mathcal{H} t}\ket{\Psi(0)}
\end{equation}
where the initial state vector $\ket{\Psi(0)}$ can be expanded on the basis of the dressed states $\{\ket{+n},\ket{-n};n=0,1,\ldots\}$ as
\begin{equation}\label{Psi_0}\begin{split}
\ket{\Psi(0)}&=\sum_{n=0}^{\infty}\Big ( \ket{+n}\braket{+n}{\Psi(0)}+\ket{-n}\braket{-n}{\Psi(0)}\Big )\\
&=\sum_{n=0}^{\infty}\Big ( c_{+n}(0)\ket{+n}+c_{-n}(0)\ket{-n}\Big )
\end{split}\end{equation}
with $c_{\pm n}(0)\equiv\braket{\pm n}{\Psi(0)}$. Inserting Eq. (\ref{Psi_0}) in Eq. (\ref{Psi_t}) one obtains:
\begin{equation}\label{Psi_t_exp}
\ket{\Psi(t)}=\sum_{n=0}^{\infty}\left ( {\rm e}^{-\frac{\imm}{\hbar}E_{+n}t}c_{+n}(0)\ket{+n}+{\rm e}^{-\frac{\imm}{\hbar}E_{-n}t}c_{-n}(0)\ket{-n}\right )
\end{equation}
where the eigenvalues $E_{\pm n}$ are given by Eq. (\ref{eigenvalues}).\\
Within a given eigenspace $\Xi(n)$, the vector state at time $t$ is
\begin{equation}
\ket{\Psi(t)}_n={\rm e}^{-\imm(n+1)\omega t}\left ( {\rm e}^{-\imm\frac{\mathcal{R}_n}{2}\omega t}c_{+n}(0)\ket{+n} + {\rm e}^{\imm\frac{\mathcal{R}_n}{2}\omega t}c_{-n}(0)\ket{-n} \right )
\end{equation}
which can be written in the interaction picture $\ket{\Psi(t)}_n\rightarrow\ket{\tilde{\Psi}(t)}_n\equiv{\rm e}^{\imm(n+1)\omega t}\ket{\Psi(t)}_n$, eliminating the free oscillation at the mean frequency $(n+1)\omega$, as
\begin{equation}\label{Psi_t_n}
\ket{\tilde{\Psi}(t)}_n=c_{+n}(t)\ket{+n}+c_{-n}(t)\ket{-n}
\end{equation}
where the evolved coefficients are defined as $c_{\pm n}(t)\equiv{\rm e}^{\mp\imm\frac{\mathcal{R}_n}{2}t}c_{\pm n}(0)$.\\
Projecting Eq. (\ref{Psi_t_n}) on the basis states $\{\bra{- n},\bra{+ n}\}$ one obtains the complex coefficients $c_{\pm n}(t)$ in the matrix form
\begin{equation}\label{c+n_t}
\begin{pmatrix}c_{- n}(t)\\c_{+ n}(t)\end{pmatrix}=
\begin{pmatrix}{\rm e}^{\imm\frac{\mathcal{R}_n}{2}t} & 0\\
0 & {\rm e}^{-\imm\frac{\mathcal{R}_n}{2}t}\end{pmatrix}
\begin{pmatrix}c_{- n}(0)\\c_{+ n}(0)\end{pmatrix}
\end{equation}
In order to discuss about the physics of the dynamics, Eq. (\ref{c+n_t}) must be expressed on the naked basis $\{\ket{e,n},\ket{g,n+1}\}$. The complex coefficients $c_{\pm n}(t)$, $c_{e,n}(t)$ and $c_{g,n+1}(t)$ are related by the unitary transformation 
\begin{equation}
U=\begin{pmatrix}\cos\theta_n & -\sin\theta_n\\
\sin\theta_n & \cos\theta_n\end{pmatrix}
\end{equation}
where $\cos\theta_n$ and $\sin\theta_n$ have been defined previously. Equation (\ref{c+n_t}) becomes
\begin{equation*}
\begin{pmatrix}c_{e,n}(t)\\c_{g,n+1}(t)\end{pmatrix}=
U^\dag\begin{pmatrix}{\rm e}^{\imm\frac{\mathcal{R}_n}{2}t} & 0\\
0 & {\rm e}^{-\imm\frac{\mathcal{R}_n}{2}t}\end{pmatrix}U
\begin{pmatrix}c_{e,n}(0)\\c_{g,n+1}(0)\end{pmatrix}
\end{equation*}
which brings to the explicit relation
\begin{equation}\label{cegn_t_general}
\begin{pmatrix}c_{e,n}(t)\\c_{g,n+1}(t)\end{pmatrix}=
\begin{pmatrix}\cos\left ( \frac{\mathcal{R}_n}{2}t\right)-\imm\frac{\delta}{\mathcal{R}_n}\sin\left ( \frac{\mathcal{R}_n}{2}t\right) & -\imm\frac{\Omega_n}{\mathcal{R}_n}\sin\left ( \frac{\mathcal{R}_n}{2}t\right) \\[1.5ex]
-\imm\frac{\Omega_n}{\mathcal{R}_n}\sin\left ( \frac{\mathcal{R}_n}{2}t\right) & \cos\left ( \frac{\mathcal{R}_n}{2}t\right)+\imm\frac{\delta}{\mathcal{R}_n}\sin\left ( \frac{\mathcal{R}_n}{2}t\right)\end{pmatrix}
\begin{pmatrix}c_{e,n}(0)\\c_{g,n+1}(0)\end{pmatrix}
\end{equation}
that describes the time evolution of the complex coefficients of the naked eigenvectors. Physically one has the complete and general description of the dynamics of the whole system, in terms of the atomic energy levels and the number of cavity photons, including detuning, coupling strength and initial conditions for pure states.

\subsection{The resonant case}
At the resonance condition $\delta=0$, the generalized Rabi frequency $\mathcal{R}_n$ coincides with the Rabi frequency $\Omega_n=2g\sqrt{n+1}$. In this case Eq. (\ref{cegn_t_general}) assumes the following simple form
\begin{equation}
\begin{pmatrix}c_{e,n}(t)\\c_{g,n+1}(t)\end{pmatrix}=
\begin{pmatrix}\cos\left ( \frac{\Omega_n}{2}t\right) & -\imm\sin\left ( \frac{\Omega_n}{2}t\right) \\[1.5ex]
-\imm\sin\left ( \frac{\Omega_n}{2}t\right) & \cos\left ( \frac{\Omega_n}{2}t\right)\end{pmatrix}
\begin{pmatrix}c_{e,n}(0)\\c_{g,n+1}(0)\end{pmatrix}
\end{equation}
As an example one can consider the initial pure state $\ket{\Psi(0)}_n=\ket{e,n}$, that is, an excited atom is injected in a cavity with the field mode in a Fock state with $n$ photons. The complex coefficients at $t=0$ are $c_{e,n}(0)=1$ and $c_{g,n+1}(0)=0$. After a certain interaction time $t>0$, the evolved coefficients are
\begin{equation}\label{cegn_t_resonant}\begin{cases}
c_{e,n}(t)&=\cos\left ( \frac{\Omega_n}{2}t \right )\\[1.5ex]
c_{g,n+1}(t)&=-\imm\sin\left ( \frac{\Omega_n}{2}t \right )
\end{cases}\end{equation}
from which the evolved state of the system can be written as
\begin{equation}\label{Initial_en}
\ket{\Psi(t)}_n=\cos\left ( \frac{\Omega_n}{2}t \right )\ket{e,n}-\imm \sin\left ( \frac{\Omega_n}{2}t \right )\ket{g,n+1}.
\end{equation}
From the relations (\ref{cegn_t_resonant}) one can calculate the probability amplitudes $P_{e,n}(t)=|c_{e,n}(t)|^2$ and $P_{g,n+1}(t)=|c_{g,n+1}(t)|^2$ corresponding, respectively, to the states $\{\ket{e,n},\ket{g,n+1}\}$\footnote{Trivially, if the initial state of the system is $\ket{\Psi(0)}=\ket{g,n+1}$, the probabilities (\ref{probs_egn}) become $P_{e,n}(t)=\sin^2\left ( \frac{\Omega_n}{2}t \right )$ and $P_{g,n+1}(t)=\cos^2\left ( \frac{\Omega_n}{2}t \right )$.}:
\begin{equation}\label{probs_egn}\begin{cases}
P_{e,n}(t)&=\cos^2\left ( \frac{\Omega_n}{2}t \right )=\frac{1}{2}\left [ 1+\cos(\Omega_nt) \right ]\\[1.5ex]
P_{g,n+1}(t)&=\sin^2\left ( \frac{\Omega_n}{2}t \right )=\frac{1}{2}\left [ 1-\cos(\Omega_nt) \right ].
\end{cases}\end{equation}
\begin{figure}[!t]
\centering
\resizebox{0.8\columnwidth}{!}{\includegraphics{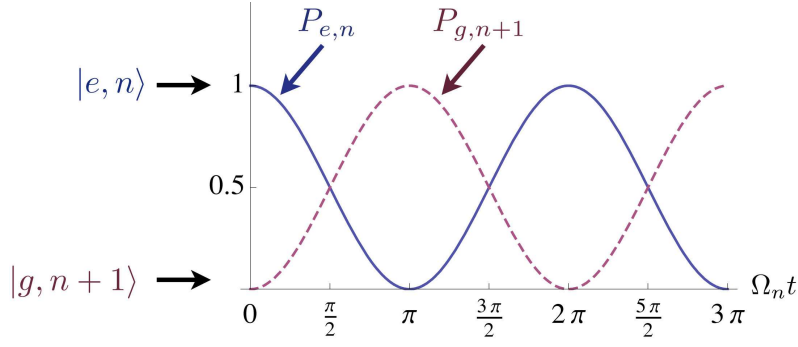}}
\caption{\label{Oscillation}Plot of Rabi oscillations for the probability amplitudes $P_{e,n}$ and $P_{g,n+1}$ in function of the dimensionless time $\Omega_nt$. The initial state of the system is $\ket{\Psi(0)}=\ket{e,n}$.}
\end{figure} 
The levels occupation probabilities of the JC system oscillate at the Rabi frequency $\Omega_n=2g\sqrt{n+1}$, frequency at which the two-level atom and the field mode exchange a single photon, process known as \emph{Rabi oscillations} or \emph{Rabi flopping}. $\Omega_n$ is, actually, the transition frequency for the dressed states, which are the energy levels to be considered in the JC model with interaction between the atom and the field mode. As plotted in Fig. \ref{Oscillation} and using Eqs. (\ref{cegn_t_resonant}), at times $\Omega_nt=k\pi$ (with $k=0,1,2\ldots$) the state of the system, if $k$ is even, is the basis eigenstate $\pm\ket{e,n}$, or $\mp\imm\ket{g,n+1}$ if $k$ is odd. At intermediate times $\Omega_nt=(2k+1)\pi/2$ the system is in an entangled state, as for instance $\frac{1}{\sqrt{2}}(\ket{e,n}-\imm\ket{g,n+1})$ when $\Omega_nt=\pi/2$.

\section{Formalization}
In this section a more mathematical treatment of the time evolution of a generic state of the system is introduced, together with the analysis of the atomic and field subsystems.

\subsection{Evolution operator}
The time evolution operator of the JC model, in interaction picture, is given by 
\begin{equation}\label{UJC}
U_{JC}(t)={\rm e}^{-\frac{\imm}{\hbar}\mathcal{H}_{AF}t}=\sum_{n=0}^\infty\frac{(-\frac{\imm}{\hbar}\mathcal{H}_{AF}t)^n}{n!}
\end {equation}
where the interaction Hamiltonian is $\mathcal{H}_{AF}=\hbar g (a r^++a^\dag r^-)$. Developing Eq. (\ref{UJC}) one can divide the sum into even and odd terms like 
\begin{equation*}
\begin{cases}
(a r^++a^\dag r^-)^{2n}=(a a^\dag)^n\ketbra{e}{e}+(a^\dag a)^n\ketbra{g}{g}\\[1.5ex]
(a r^++a^\dag r^-)^{2n+1}=(a a^\dag)^n a\ketbra{e}{g}+a^\dag (a a^\dag)^n\ketbra{g}{e}
\end{cases}
\end{equation*}
resulting in a time evolution operator of the form
\begin{equation}\begin{split}
U_{JC}(t)=\sum_{n=0}^\infty&\Big \{ \frac{(-\imm g t)^{2n}}{(2n)!}\left [ (\sqrt{a a^\dag})^{2n} \ketbra{e}{e}+ (\sqrt{a^\dag a})^{2n} \ketbra{g}{g}\right ]+ \\
 &+\frac{(-\imm g t)^{2n+1}}{(2n+1)!}\left [ \frac{(\sqrt{a a^\dag})^{2n+1}}{\sqrt{aa^\dag}} a\ketbra{e}{g}+ a^\dag\frac{(\sqrt{a a^\dag})^{2n+1}}{\sqrt{aa^\dag}}\ketbra{g}{e}\right ]\Big \}
\end{split}\end{equation}
that, rewritten in a more compact way, it appears as
\begin{equation}\begin{split}
U_{JC}(t)&= \cos(gt\sqrt{N+1})\ketbra{e}{e}+ \cos(gt\sqrt{N}) \ketbra{g}{g}+ \\
 &-\imm\frac{\sin(gt\sqrt{N+1})}{\sqrt{N+1}} a\ketbra{e}{g}-\imm a^\dag\frac{\sin(gt\sqrt{N+1}))}{\sqrt{N+1}}\ketbra{g}{e}
\end{split}\end{equation}
where the number of photons operator is $N=a^\dag a$. Written in a matrix form the time evolution operator assumes the form
\begin{equation}
U_{JC}(t)=\begin{pmatrix} u_{gg}(t) & u_{ge}(t) \\ u_{eg}(t) & u_{ee}(t) \end{pmatrix}
\end{equation}
where $ u_{gg}(t)=\cos(gt\sqrt{N})$, $ u_{ge}(t)=-\imm a^\dag\frac{\sin(gt\sqrt{N+1}))}{\sqrt{N+1}}$, $ u_{eg}(t)=-\imm\frac{\sin(gt\sqrt{N+1})}{\sqrt{N+1}} a$ and $ u_{ee}(t)=\cos(gt\sqrt{N+1})$.

\subsubsection{Pure state}
A generic initial pure state for the atom-field system can be written on the naked basis $\{\ket{e,n},\ket{g,n}\}$ (which includes also the ground state $\ket{g,0}$) in the form
\begin{equation}
\ket{\Psi(0)}=\sum_{n=0}^\infty(c_{e,n}(0)\ket{e,n}+c_{g,n}(0)\ket{g,n})
\end{equation}
with $c_{e,n}(0)=c_e(0)c_n(0)$ and $c_{g,n}(0)=c_g(0)c_n(0)$. The evolved state $\ket{\Psi(t)}=U_{JC}\ket{\Psi(0)}$ is
\begin{equation}\begin{split}\label{pure_gen}
\ket{\Psi(t)}=&\sum_{n=0}^\infty c_n(0)\Big \{ c_e(0)\left [ \cos\left ( \frac{\Omega_n}{2}t\right ) \ket{e,n}-\imm\sin\left ( \frac{\Omega_n}{2}t\right ) \ket{g,n+1} \right ]+\\
&+c_g(0)\left [ \cos\left ( \frac{\Omega_{n-1}}{2}t\right ) \ket{g,n}-\imm\sin\left ( \frac{\Omega_{n-1}}{2}t\right ) \ket{e,n-1} \right ]\Big \}.
\end{split}\end{equation}
As an example one could take as the initial state of the system $\ket{\Psi(0)}=\ket{e,n}$, verifying that the evolved state (\ref{pure_gen}) is exactly that of Eq. (\ref{Initial_en}).

\subsubsection{Mixed state}
A generic initial mixed state of the system is described by the density operator $\rho(0)=\rho_{F}(0)\otimes\rho_{A}(0)$, where the field and atomic subsystems can be considered factorized at $t=0$. The time evolution of the system is 
\begin{equation}
\rho(t)=U_{JC}(t)\rho(0)U_{JC}^\dag(t).
\end{equation}
As an example on could take as initial state of the system the density operator $\rho(0)=\rho_F(0)\otimes\ketbra{e}{e}$, where the atom is in the excited state and the cavity mode in a generic mixed state, and the evolved state of the system is
\begin{equation}\label{rhot_ee}\begin{split}
\rho(t)&=u_{ee}(t)\rho_F(0)u_{ee}^\dag(t)\ketbra{e}{e}+u_{ge}(t)\rho_F(0)u_{ge}^\dag(t)\ketbra{g}{g}+\\
&+u_{ee}(t)\rho_F(0)u_{ge}^\dag(t)\ketbra{e}{g}+u_{ge}(t)\rho_F(0)u_{ee}^\dag(t)\ketbra{g}{e}.
\end{split}\end{equation}

\subsection{Atomic time evolution}
In order to consider only the atomic subsytem dynamics, one must apply a partial trace of the system density operator over the field degrees of freedom. Starting, as usual, from the simple case of $\rho(0)=\rho_F(0)\otimes\ketbra{e}{e}$ and using Eq. (\ref{rhot_ee}) the density operator describing the atomic time evolution is 
\begin{equation}\label{rho_A_general}\begin{split}
\rho_A(t)={\rm Tr}_F[\rho(t)]&={\rm Tr}_F[u_{ee}(t)\rho_F(0)u_{ee}^\dag(t)]\ketbra{e}{e}+{\rm Tr}_F[u_{ge}(t)\rho_F(0)u_{ge}^\dag(t)]\ketbra{g}{g}\\
&+{\rm Tr}_F[u_{ee}(t)\rho_F(0)u_{ge}^\dag(t)]\ketbra{e}{g}+{\rm Tr}_F[u_{ge}(t)\rho_F(0)u_{ee}^\dag(t)]\ketbra{g}{e}.
\end{split}\end{equation}
For an experimental measurement the observables are the atomic probabilities, which can be expressed as
\begin{equation}\label{PePg}
\begin{cases}
P_e(t)=\meanvalue{e}{\rho_A(t)}{e}={\rm Tr}_F[u_{ee}^\dag(t)u_{ee}(t)\rho_F(0)]=\langle u_{ee}^\dag u_{ee}\rangle(t)\\[1.5ex]
P_g(t)=\meanvalue{g}{\rho_A(t)}{g}={\rm Tr}_F[u_{ge}^\dag(t)u_{ge}(t)\rho_F(0)]=\langle u_{ge}^\dag u_{ge}\rangle(t)
\end{cases}
\end{equation}
where the cyclic property of the trace operation\footnote{The cyclic property of the trace states: ${\rm Tr}[ABC]={\rm Tr}[CAB]={\rm Tr}[BCA]$.} has been used. As can be easily veryfied, the sum of the probabilities must give one, $$P_e(t)+P_g(t)={\rm Tr}_F[(u_{ee}^\dag(t)u_{ee}(t)+u_{ge}^\dag(t)u_{ge}(t))\rho_F(0)]={\rm Tr}_F[\mathbf{I}_F\rho_F(0)]=1$$
A particular example can be considered, that of the field mode prepared in a diagonal form $\rho_F(0)=\sum_{m}P_m\ketbra{m}{m}$, with the normalization condition $\sum_{m}P_m=1$ and where $P_m$ is the probability corresponding to the Fock state $\ket{m}$. This is the case of the field in a Fock state $\ket{n}$ with $P_m=\delta_{m,n}$, or in a thermal state with $P_m=\frac{1}{1+\langle N\rangle}\left ( \frac{\langle N\rangle}{1+\langle N\rangle} \right)^m$, where $\langle N\rangle$ is the average number of thermal photons. In this case, the off diagonal terms in Eq. (\ref{rho_A_general}) are null and the density operator for the atomic subsystem is a statistical mixture of the atomic states $\ket{e}$ and $\ket{g}$:
\begin{equation}
\rho_A(t)=P_e(t)\ketbra{e}{e}+P_g(t)\ketbra{g}{g}
\end{equation}
where the probabilities $P_e(t)$ and $P_g(t)$ are calculated in Eqs. (\ref{PePg}).

\subsection{Field mode time evolution}
The dynamics of the field mode may occur in two ways, depending on  
whether the state of the atom exiting the cavity is measured or not. In general the two types of measurement are labelled as ``selective'' and ``non-selective'' measurements which are, respectively, a non-unitary and a unitary evolution.\\
If, at a particular time $t$, the atom is measured outside the cavity it has passed through, the field mode inside the cavity, which is entangled with the atom due to their previous interaction, undergoes a non-unitary evolution. The field state is projected into a normalized state, quantum correlated with the atomic state measured at that particular time. As before, one can consider as the initial state of the system $\rho(0)=\rho_F(0)\otimes\ketbra{e}{e}$ and, depending on the measured atomic state $\ket{e}$ or $\ket{g}$, the field state is 
\begin{equation}\label{field_atom_meas}
\rho_F(t)=\begin{cases}\rho_F^{(e)}(t)=\frac{u_{ee}(t)\rho_F(0)u_{ee}^\dag(t)}{{\rm Tr}_F[u_{ee}(t)\rho_F(0)u_{ee}^\dag(t)]}=\frac{u_{ee}(t)\rho_F(0)u_{ee}^\dag(t)}{P_{e}(t)}\quad\text{atom measured in }\ket{e}\\[2ex]
\rho_F^{(g)}(t)=\frac{u_{ge}(t)\rho_F(0)u_{ge}^\dag(t)}{{\rm Tr}_F[u_{ge}(t)\rho_F(0)u_{ge}^\dag(t)]}=\frac{u_{ge}(t)\rho_F(0)u_{ge}^\dag(t)}{P_{g}(t)}\quad\text{atom measured in }\ket{g}
\end{cases}
\end{equation}
On the other hand, if the atom is not measured or, equivalently, the outcome concerning the atomic state is ignored (non-selective measurement), the field state will be a statistical mixture of the two possible outcomes (\ref{field_atom_meas}):
\begin{equation}\label{field_atom_NOmeas}\begin{split}
\rho_F(t)&=P_e(t)\rho_F^{(e)}(t)+P_g(t)\rho_F^{(g)}(t)=u_{ee}(t)\rho_F(0)u_{ee}^\dag(t)+u_{ge}(t)\rho_F(0)u_{ge}^\dag(t)\\
&=\mathcal{F}^{(e)}_{\text{JC}}(t)\rho_F(0)
\end{split}\end{equation}
where $\mathcal{F}^{(e)}_{\text{JC}}(t)$ is the ``super-operator'' of Jaynes-Cummings for the reduced dynamics of the field alone, with the atom initially in the excited state. This is also the way of writing a \emph{quantum map}, that is an operation that transform a density operator in another density operator, leading to a physically acceptable description of the system evolution \cite{Nielsen_Chuang}. Usually it is written as a \emph{Kraus sum}
\begin{equation}
\rho\rightarrow\mathcal{L}(\rho)=\sum_{\mu}M_\mu\rho M_\mu^\dag,\qquad\text{with }\sum_\mu M^\dag_\mu M_\mu=\mathbf{I}
\end{equation}
where $M_\mu$ are called \emph{Kraus operators}. As regards the system above, the Kraus operators are $\{u_{ee}(t),u_{ge}(t)\}$, while $F_e\equiv u_{ee}^\dag(t)u_{ee}(t)$ and $F_g\equiv  u_{ge}^\dag(t)u_{ge}(t)$ are called POVM (Positive Operator Valued Measure), which are generalized measurement operators that include the class of projectors $P_j$ with the property $P_j^\dag=P_j$. POVMs must satisfy only the property of summing up to one, that is $F_e+F_g=\mathbf{I}$, which ensures to have positive probabilities as $P_{e;g}={\rm Tr}[\rho F_{e;g}]$.

\section{Vacuum Rabi oscillations}
Retrieving the previous analysis on the Rabi oscillations for resonant condition $\delta=0$, the particular case with the initial condition $\ket{\Psi(0)}=\ket{e,0}$ can be considered. In this case the cavity is in the vacuum state with zero photons ($n=0$), while the atom is prepared in its excited state $\ket{e}$. The probabilities for the naked basis states $\{\ket{e,0},\ket{g,1}\}$ are given by Eqs. (\ref{probs_egn}) with $n=0$ resulting in
\begin{equation}\begin{cases}
P_{e,0}(t)=\cos^2\left ( \frac{\Omega_0}{2} t \right)=\frac{1}{2}\left [ 1+\cos(\Omega_0t) \right ]\\[1.5ex]
P_{g,1}(t)=\sin^2\left ( \frac{\Omega_0}{2} t \right)=\frac{1}{2}\left [ 1-\cos(\Omega_0t) \right ]
\end{cases}\end{equation}
where $\Omega_0=2g$ is the \emph{vacuum Rabi frequency}, that is the transition frequency for the dressed states $\ket{\pm 0}=(\ket{e,0}\pm\ket{g,1})/\sqrt{2}$ in the subspace $\Xi(0)$, with energies 
\begin{equation}
E_{\pm 0}=\hbar(\omega\pm\frac{\Omega_0}{2})=\hbar(\omega\pm g).
\end{equation}
The energy separation between the two dressed levels is $(\Delta E)_0=2\hbar g$ and this particular situation is known as \emph{vacuum Rabi oscillations} or \emph{1-photon oscillations}. This means that the spontaneous emission of a photon by the atom inside a cavity becomes a reversible process, with an oscillatory behavior in time, which is in hard contrast with the irreversible spontaneous emission in free space, with exponential decay in time. This is due to the radically different energy spectral distribution of the electromagnetic field in the two cases: in the first there is only one field mode in the cavity accessible by the emitted photon, in the second there is a continuum of available modes. On the other hand, the spontaneous emission in a cavity can be suppressed if there is not any field mode resonant with the atomic transition. Modifying the field spectral distribution the spontaneous emission can be enhanced or suppressed\footnote{The spontaneous emission can be also suppressed in dielectric media with periodically time changing refractive index (photonic crystals), where there are energy bands ``forbidden'' for photons, just like the case of electronic bands in crystals.} (cavity QED).\\
A simple explanation of the vacuum Rabi oscillatons can be achieved by thinking the atom (weakly excited since the field mode has a maximum of one photon) as an oscillating dipole, in particular a quantized harmonic oscillator. The interacting atom and mode can be thought as two weakly coupled oscillators at the same frequency, described by the Hamiltonian
\begin{equation}
\mathcal{H}_{JC}\rightarrow\mathcal{H}=\hbar\omega(b^\dag b+a^\dag a)+\hbar g (a^\dag b+ab^\dag)
\end{equation}
where the commutation relations holds for the two subsystems $[a,a^\dag]=1$ (mode) and $[b^\dag,b]=1$ (atom). This Hamiltonian can be diagonalized introducing the two normal modes operators
\begin{equation}
c\equiv\frac{a+b}{\sqrt{2}}\qquad,\qquad d\equiv\frac{a-b}{\sqrt{2}}
\end{equation}
obtaining the new one
\begin{equation}
\mathcal{H}=\hbar(\omega+g)c^\dag c+\hbar(\omega-g)d^\dag d.
\end{equation}
This is the Hamiltonian describing two independent harmonic oscillators (normal modes) of different frequencies $\omega\pm g$.\\
The eigenstates of $\mathcal{H}$ are
\begin{equation}
\mathcal{H}\ket{n_c,n_d}=[\hbar(\omega+g)n_c+\hbar(\omega-g)n_d]\ket{n_c,n_d}
\end{equation}
which become for the lowest states
\begin{equation}
\mathcal{H}\ket{1_c,0_d}=\hbar(\omega+g)\ket{1_c,0_d}\quad;\quad\mathcal{H}\ket{0_c,1_d}=\hbar(\omega-g)\ket{0_c,1_d}.
\end{equation}
In terms of the original mode operators $a$ and $b$, the eigenstates can be compared to the dressed states for the atom and the cavity field mode:
\begin{equation}\begin{cases}
\ket{1_c,0_d}=\frac{1}{\sqrt{2}}\left ( \ket{1_a,0_b}+\ket{0_a,1_b} \right )\\[1.5ex]
\ket{0_c,1_d}=\frac{1}{\sqrt{2}}\left ( \ket{1_a,0_b}-\ket{0_a,1_b} \right )
\end{cases}
\longleftrightarrow\quad\begin{cases}
\ket{+0}=\frac{1}{\sqrt{2}}\left ( \ket{e,0}+\ket{g,1} \right )\\[1.5ex]
\ket{-0}=\frac{1}{\sqrt{2}}\left ( \ket{e,0}-\ket{g,1} \right )
\end{cases}
\end{equation}
Hence, the lowest stationary states of two harmonic oscillators correspond to the energies $\hbar(\omega\pm g)$, exactly as the first two dressed states $\ket{\pm 0}$ of the lowest manifold $\Xi(0)$ of the JC model.\\
Experimental confirmations of Rabi oscilations were given by the group of Kimble \cite{Kimble}, in the optical domain, and by the group of Haroche \cite{Brune_Haroche}, in th microwave domain. The atom-field system in the vacuum state can be weakly excited by a probe field tunable in frequency. The outcome is that the excitation is not resonant with the field mode (or the atomic) frequency $\omega$, but, as predicted, with the two frequencies of the normal modes ($c$ and $d$) of the coupled system: $\omega\pm g$.

\subsection{An application: swapping of the excitation}\label{Swapping}
Suppose that a two-level atom, prepared in the superposition state $c_e\ket{e}+c_g\ket{g}$, is injected in a cavity in the vacuum state $\ket{0}$, giving as initial state for the whole system
\begin{equation}\label{Initial_swapping}
\ket{\Psi(0)}=(c_e\ket{e}+c_g\ket{g})\ket{0}.
\end{equation}
Using Eqs. (\ref{pure_gen}), one derives the following simple time evolutions for the states $\ket{e,0}$ and $\ket{g,1}$:
\begin{equation}\label{Eq_swapping}\begin{cases}
\ket{e,0}\rightarrow \cos\left ( \frac{\Omega_0}{2}t\right ) \ket{e,0}-\imm\sin\left ( \frac{\Omega_0}{2}t\right ) \ket{g,1} \\[1.5ex]
\ket{g,1}\rightarrow \cos\left ( \frac{\Omega_{0}}{2}t\right ) \ket{g,1}-\imm\sin\left ( \frac{\Omega_{0}}{2}t\right ) \ket{e,0}\\[1.5ex]
\ket{g,0}\rightarrow\ket{g,0}\quad\text{(uncoupled)}
\end{cases}\end{equation}
If the system evolves for a time $t$ such that $\Omega_0t=\pi$ the above system of equations give the results
\begin{equation}\begin{cases}
\ket{e,0}\rightarrow-\imm\ket{g,1}\\[1.5ex]
\ket{g,1}\rightarrow-\imm\ket{e,0}\\[1.5ex]
\ket{g,0}\rightarrow\ket{g,0}
\end{cases}\end{equation}
and the whole system starting from the state (\ref{Initial_swapping}) is in the state
\begin{equation}\label{Evolved_swapping}
\ket{\Psi(t)}=\ket{g}(-\imm c_e\ket{1}+c_g\ket{0})
\end{equation}
from which it is evident that the initial superposition state of the atom has been swapped with (or transferred to) the state of the cavity mode (except for a local phase factor).\\
Since the process is reversible, one may imagine that after the time $t$ the atom goes outside the cavity and another one is injected in the ground state $\ket{g}_2$ to interact with the cavity mode (now the state of the whole system is given by Eq. (\ref{Evolved_swapping})). After a second time evolution with $\Omega_0 t=\pi$ the outcoming state of the system is
\begin{equation}
\ket{g}_2(-\imm c_e\ket{1}+c_g\ket{0})\rightarrow(-c_e\ket{e}_2+c_g\ket{g}_2)\ket{0}
\end{equation}
which means that the photonic state of the cavity now has been swapped with the second atom, recovering a state similar to the initial one.

\section{Collapses and revivals}
Going back to the general case of $n\neq 0$ (always for resonance conditions), it has been shown that if the field mode is prepared in a Fock state $\ket{n}$ and the atom injected in the cavity in the state $\ket{e}$, the probability to find the atom again in its excited state after a time $t>0$ is given by the first of Eqs. (\ref{probs_egn}).\\
If the field state is prepared in a generic superposition of Fock states $\rho_F(0)=\sum_{n=0}^\infty P_n\ketbra{n}{n}$, the probability to find the atom in the excited state is
\begin{equation}
P_{e}(t)=\sum_{n=0}^\infty P_n P_{e,n}(t)=\frac{1}{2}\left [ 1+\sum_{n=0}^\infty P_n\cos\left ( \frac{\Omega_n t}{2}\right )\right ]
\end{equation}
which is an average over all the possible (infinite) Fock states $\ket{n}$ weighed by the probabilities $P_n$.\\
\begin{figure}[!h]
\centering
\resizebox{0.55\columnwidth}{!}{\includegraphics{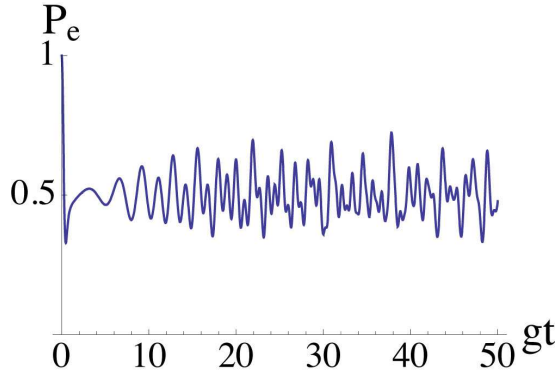}}
\caption{\label{ColRev_thermal}Collapse of Rabi oscillations for the cavity field initially in a thermal state with $\langle N\rangle=10$ and the atom in the excited state $\ket{e}$.}
\end{figure} 
If, in particular, the initial state of the cavity is a thermal state with $P^{(T)}_n=\frac{\langle N\rangle^n}{(1+\langle N\rangle)^{n+1}}$, the infinite sum of sinusoidal oscillations, each weighed by $P_n$, brings the state to a rapid collapse of Rabi oscillations, until $P_e(t)\rightarrow\frac{1}{2}$ with some small fluctuations (see Fig. \ref{ColRev_thermal}). In this way the probabilty to observe the atom in its excited or ground state is equally distributed, exhibiting a ``classical'' effect.\\
\begin{figure}[!h]
\centering
\resizebox{0.55\columnwidth}{!}{\includegraphics{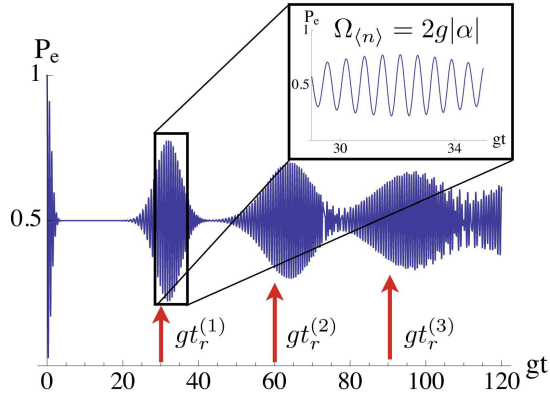}}
\caption{\label{Revivals}Collapses and revivals of Rabi oscillations for the cavity field initially in a coherent state $\ket{\alpha}$ with $|\alpha|^2=25$ and the atom in the excited state $\ket{e}$. In evidence there are three of the revival times $gt_r^{(j)}=j2\pi|\alpha|$ and in the inset it is shown the Rabi oscillation at the central frequency $\Omega_{\langle n\rangle}$.}
\end{figure} 
Different is the case of an initial coherent state $\ket{\alpha}$ of the cavity field. If the coherent state amplitude is too high, the field is ``classical'' and, as for the Fock state $\ket{n}$, there is ideally a Rabi oscillation at a mean frequency $\Omega_{\langle n \rangle}\simeq 2g|\alpha|=2g\sqrt{\langle n \rangle_\alpha}$ (semi-classical thoery, see Ramsey cavities). For a coherent state $P^{(C)}_n={\rm e}^{-|\alpha|^2}\frac{|\alpha|^{2n}}{n!}$ and the atomic excitation probability is
\begin{equation}
P_e(t)=\frac{1}{2}\left [ 1+{\rm e}^{-|\alpha|^2}\sum_{n=0}^\infty \frac{|\alpha|^{2n}}{n!}\cos(\Omega_n t)\right ].
\end{equation}
First of all it is possible to demonstrate that for $|\alpha|^2\gg 1$ the probability for the atomic excited state is
\begin{equation}
P_e(t)\simeq\frac{1}{2}\left [ 1+ \cos(2 g|\alpha| t){\rm e}^{-\frac{g^2t^2}{2}}\right ]
\end{equation}
where the oscillations occur at the mean Rabi frequency $\Omega_{\langle n\rangle}=2g\sqrt{\langle n\rangle_{\alpha}}=2g|\alpha|$ with a Gaussian envelope of width $t_c\sim\frac{1}{g}$, known also as \emph{coherence time} (or collapse time). This result can be interpreted in terms of a destructive interference of Rabi oscillations at different frequencies, where only a narrow range of Rabi frequencies $\Delta\Omega\approx g$ around $\Omega_{\langle n\rangle}$ plays a role in the dynamics (in contrast with the thermal case). The mechanism of collapse is always interpretable as a ``classical'' effect.\\
What is a purely quantum feature is the \emph{revival effect} of Rabi oscillations, which occurs with a period of time $$t_r\simeq 2\pi\frac{|\alpha|}{g}>t_c.$$
The revival time is the period of time for which two Rabi oscillations at two close frequencies (and necessarily close to the central frequency) provide a constructive interference. This happens when the two oscillations get in phase and an extimate of the revival time is the following
\begin{equation}
t_r^{(j)}=\frac{2\pi j}{\Omega_{\langle n\rangle}-\Omega_{\langle n\rangle-1}}=\frac{2\pi j}{2g(|\alpha|-\sqrt{|\alpha|^2-1})}\simeq j\frac{2\pi |\alpha|}{g}
\end{equation}
for $|\alpha|^2\gg 1$, but not too high.
The revival effect, shown in Fig. \ref{Revivals} is a phenomenon of quantum recurrence due to the ``granular'' nature of the quantized electromagnetic field and it allows the oscillations at different frequencies to be periodically in phase. For long times the effect gets more and more ruined since the revivals are never complete.

\section{The Haroche experiment}
In 1996 the group of Haroche, from the Ecole Normale Sup\'erieure of Paris, published a paper \cite{Brune_Haroche} named ``Quantum Rabi Oscillation: A Direct Test of Field Quantization in a Cavity''. The realization of the experiment required a resonant interaction between a two-level atom and one cavity mode, a strong coupling between the two subsystems with an interaction time shorter than the atomic decay, a high-Q cavity, the control of paramaters like atomic excitation and speed, and cavity temperature. A superconductive microwave cavity was chosen, through which the atoms, selected in velocity and excited at Rydberg levels (main quantum number $n\gg 1$), passed one at a time, with one Rydberg transition resonant with the cavity mode.\\
\begin{figure}[!h]
\centering
\resizebox{0.75\columnwidth}{!}{\includegraphics{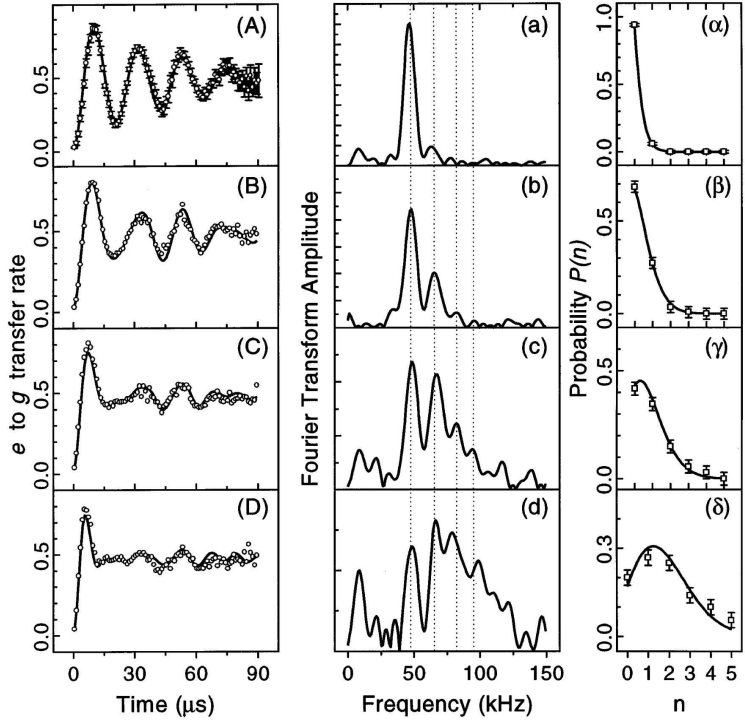}}
\caption{\label{HarocheExp} Results of the Haroche experiment \cite{Brune_Haroche}, which put in evidence collapses and revivals of Rabi oscillations, depending on the mean photon number present in the cavity: (A) $\langle n \rangle_{\alpha}\simeq 0$, (B) $\langle n \rangle_{\alpha}\simeq 0.40$, (C) $\langle n \rangle_{\alpha}\simeq 0.85$ and (D) $\langle n \rangle_{\alpha}\simeq 1.77$. (a), (b), (c), (d) are the corresponding Fourier transforms, whereas ($\alpha$), ($\beta$), ($\gamma$), ($\delta$) are the corresponding photon number distributions.}
\end{figure}
In particular the group of researchers used Rubidium atoms excited in circular Rydberg states ($\ket{e}\leftrightarrow n_e=51$ and $\ket{g}\leftrightarrow n_g=50$), with a transition frequency between the two states $\nu_{eg}\simeq 51$GHz. The average lifetime of the atomic excited level was $\gamma_A^{-1}\simeq 32$ms, whereas the interaction time between the atom and the cavity mode was $t_{\rm int}\simeq 10$ms. Using Rydberg atoms ensures that the dipole coupling ($d\propto n^2$) is very strong, more precisely the coupling constant was $g\simeq 1.6\times10^5$Hz. The superconductive microwave cavity ($L\simeq 3$cm) supported one mode resonant with di Rydberg transition $\nu=\nu_{eg}$. The cavity was cooled at a teperature of $T=0.8$K, corresponding to $\bar{n}\simeq 0.05$ thermal photons. The quality factor of the cavity was $Q=\frac{\omega}{2\kappa}\simeq 108$, which means that the average lifetime of cavity photons was $\frac{1}{2\kappa}\simeq 220\mu$s. The values of these parameters allowed a \emph{strong coupling regime}, for which the coupling frequency was much greater than all the decay rates $g\gg \{2\kappa,\gamma_A\}$ and a coherent JC dynamics may occur. Moreover a source of microwave radiation could inject into the cavity a small coherent field (from zero to few photons) with a high controlled energy. Detectors based on field ionization techniques allowed to count the outcoming atoms in the excited or in the ground state. The atoms were prepared, injected and measured one at a time, separated by a reset time ($=2.5$ms). It was possible to obtain the occupation probabilities $P_e$ and $P_g$ in function of the interaction time $t_{\rm int}$, which could be changed varying the atomic velocities. After reaching a sufficient statistics, the ensemble of measurements was repeated for different interaction time values. These measurements referred to the probability 
\begin{equation*}
P_g(t)=1-P_e(t)=\frac{1}{2}\left [ 1-\sum_{n}P_n\cos(\Omega_n t)\right ]
\end{equation*}
with $t\equiv t_{\rm int}$ (Fig. \ref{HarocheExp}A-D), and also to the Fourier transform of $P_g(t)$ (Fig. \ref{HarocheExp}a-d) and to photon statistics (Fig. \ref{HarocheExp}$\alpha$-$\delta$)\footnote{Reprinted figure with permission from  M. Brune \textit{et al.}, Phys. Rev. Lett. \textbf{76}, 1800 (1996). Copyright (1996) by the American Physical Society. http://link.aps.org.pros.lib.unimi.it/abstract/PRL/v76/p1800 }. All these measured quantities were repeated for different coherent field amplitudes of the cavity mode state. In the first case $\langle n\rangle_{\alpha}=0$, the vacuum Rabi oscillations were observed, for the first time in the time domain, at the frequency $\nu_0=\frac{\Omega_0}{2\pi}\simeq47$kHz. For a greater coherent field amplitude inside the cavity $\langle n\rangle_{\alpha}=0.4, 0.85, 1.77$ the signal $P_g(t)$ lost progressively its sinusoidal behavior, with an evidence of the presence of collapses and revivals in Fig. \ref{HarocheExp}C-D. From the Fourier transforms plots (a-d) there is evidence of subsequent Rabi frequencies involved in the dynamics, as the coherent field amplitude increases, in particular the frequencies $\nu_0\sqrt(2),\nu_0\sqrt(3),2\nu_0$ as predicted by the theory $\nu_n=\frac{\Omega_n}{2\pi}=\nu_0\sqrt{n+1}$. This is the witness of the granular nature of the quantized electromagnetic field inside the cavity. The weights of the Fourier components yielded the photon number distributions in the field ($\alpha$-$\delta$). For no injected field the pionts are in good agreement with the theoretical exponential decay curve of the thermal radiation. When a coeherent field is present in the cavity, the photon number distribution follows a Poisson statistics, signature of a coherent radiation.

\section{Generation of two atoms entangled states}
Another important experiment from the group of Haroche \cite{Hagley_Haroche}, published in 1997 with a paper named ``Generation of Einstein-Podolsky-Rosen Pairs of Atoms'', based essentially on the same apparutus as the previous one, exploited the resonant JC interaction to generate entangled states of two atoms via a cavity mode.\\
At time $t=0$ a first circular Rydberg two-level atom was prepared in the excited state $\ket{e}_1$ and injected in a superconductive microwave cavity in the vacuum state $\ket{0}$: $$\ket{\Psi(0)}=\ket{e}_1\otimes\ket{0}\equiv\ket{e_1,0}.$$
After a certain interaction time $t_1$ such that $\Omega_0 t_1=\frac{\pi}{2}$, where $\Omega_0=2g$ is the vacuum Rabi frequency, the first atom interacted with the cavity field mode. Remembering Eqs. (\ref{Eq_swapping}) in Section \ref{Swapping} and using real coefficients, one has
\begin{equation}\begin{cases}
\ket{e,0}\rightarrow \cos\left ( \frac{\Omega_0}{2}t\right ) \ket{e,0}+\sin\left ( \frac{\Omega_0}{2}t\right ) \ket{g,1} \\[1.5ex]
\ket{g,1}\rightarrow \cos\left ( \frac{\Omega_{0}}{2}t\right ) \ket{g,1}+\sin\left ( \frac{\Omega_{0}}{2}t\right ) \ket{e,0}
\end{cases}\end{equation}
which brings to evolved state 
\begin{equation}
\ket{e_1,0}\rightarrow\ket{\Psi(t_1)}=\frac{1}{\sqrt{2}}\left ( \ket{e_1,0}+\ket{g_1,1} \right ).
\end{equation}
After a time interval $T$ a second atom, prepared now in the ground state $\ket{g_2}$, was injected in the cavity and the state is $$\ket{\Psi(t_1+T)}=\ket{g_2}\otimes\frac{1}{\sqrt{2}}\left ( \ket{e_1,0}+\ket{g_1,1} \right )=\frac{\ket{e_1,g_2,0}+\ket{g_1,g_2,1}}{\sqrt{2}}.$$
The second atom interacted with the cavity mode for a time $t_2=2t_1$, such that $\Omega_0 t_2=\pi$ and, since the state $\ket{g_2,0}$ is uncoupled, the state of the system was
$$\ket{\Psi(t_1+T+t_2)}=\frac{\ket{e_1,g_2,0}+\ket{g_1,e_2,0}}{\sqrt{2}}=\ket{\Psi^-}_{A1A2}\otimes\ket{0}_F$$
where $\ket{\Psi^-}_{A1A2}=\frac{\ket{e_1,g_2}+\ket{g_1,e_2}}{\sqrt{2}}$ is one of the maximally entangled states of the Bell basis. The cavity field was in the vacuum state and it ``catalysed'' the process of entangling two independent atoms through the common and subsequent JC interaction. After a time $t>t_1+T+t_2$ the energy of the two atoms was measured or, in order to confirm the generation of the entangled state, interference effects measurements (Ramsey fringes) or correlation measurements (Bell signal) were performed.

\section{Conclusions}
This tutorial intends to provide the basic elements for the comprehension of the Jaynes-Cummings model, in order to give a pedagogical approach to the study of matter-radiation interaction with the tools of quantum mechanics. Along with this approach, the Hamiltonian of the interacting system is derived from general statements and its eigenvalues and eigenvectors are calculated from diagonalization. The power and simplicity of the JC model stem from the fact that the dynamics occurs within a bidimensional subspace spanned by the dressed basis states, which are related to the naked basis states by a simple rotation matrix. The description of the dynamics by means of these two basis explains in an elegant way the Rabi oscillations phenomenon, that is the coherent exchange of energy between a two-level system and a resonant electromagnetic field mode. A particular attention is devoted to the vacuum Rabi oscillations, where the initial state of the field mode contains zero photons.
The formal description of the evolution operator is presented in order to obtain the JC dynamics for every initial state of the system, including the time evolution of the subsystems. The real quantum feature of the JC model is given by the revival effect, occurring with an initial coherent state of the field mode, which can be explained only if the electromagnetic field is quantized.\\
cQED is a natural framework for the implementation of the JC model, where single two-level Rydberg atoms can pass through or be trapped in microwave single mode cavities. Two key-experiments are described in the last sections, where Rabi oscillations, together with their collapse and revival effects, and the swapping application for retrieving maximally entangled states, are the results of spectacular observations and measurements.

\end{document}